\documentstyle[12pt]{article}
\begin{document}
\def\lsim{\ \matrix{<\cr\noalign{\vskip-7pt}\sim\cr} \ }
\def\gsim{\ \matrix{>\cr\noalign{\vskip-7pt}\sim\cr} \ }
\begin{titlepage}
\hfill{PURD-TH-93-04}
\vspace{5 mm}

\begin{center}
{\Large {\bf A Shadow Matter Universe}}\\
\vspace{7 mm}
H. Lew\footnote{email: lew\%purdd.hepnet@LBL.Gov} \\
\vspace{3 mm}
{\it Physics Department,\\ Purdue University,\\ West Lafayette,
IN47907-1396, U.S.A.}
\end{center}

\vspace{8 mm}
\centerline{ABSTRACT}
\vspace{2 mm}

The cosmological and astrophysical implications of a shadow
matter model which could also have interesting experimental
consequences are examined. The model has identical microphysics
for both the ordinary and shadow worlds but requires a
macroscopic asymmetry from nucleosynthesis constraints.
It is proposed that this macroscopic asymmetry can be generated
at the quark-hadron phase transition.
\vfill
\leftline{PACS number(s): 98.80.Cq, 12.15.Cc}
\vspace{2 mm}
\end{titlepage}

{}From time to time there have been some speculation about the
existence of ``shadow matter'' in the Universe and its
consequences \cite{mp,kst,flv}. This shadow matter is another
form of matter which is thought to interact with ordinary matter
via gravity or via interactions of comparable strength to gravity.
Some of the astrophysical and cosmological implications of generic
shadow matter models have been studied in Ref.\cite{kst}. In this
paper the astrophysical and cosmological consequences of a particular
realization of the shadow matter scenario, which may also have
interesting collider physics, will be discussed.

The construction of a shadow matter model is quite straightforward.
Consider the Lagrangian
\begin{equation}
{\cal L} = {\cal L}_1 + {\cal L}_2
\end{equation}
where ${\cal L}_{1,2}$ describes the physics of the ordinary
and shadow worlds respectively. If the requirement that the
microphysics of the two worlds be identical is imposed then
there exists a discrete symmetry such that
\begin{equation}
{\cal L}_1 \longleftrightarrow {\cal L}_2.
\end{equation}
Let $G_{1,2}$ be the gauge group of ${\cal L}_{1,2}$ where
$G_1$ and $G_2$ are isomorphic and interchange under the
discrete symmetry. The gauge group of ${\cal L}$ is then
$G_1 \otimes G_2$ with the particle content of the ordinary
and mirror worlds described by the representations
$(R,1) \oplus (1,R)$. For example, the gauge group could
be $E_8 \otimes E_8$ as motivated by superstring theories.
The discrete symmetry between the ordinary and shadow worlds
may or may not survive at low energies.

In the following the gauge group, $G_{1,2}$, is taken to be
the Minimal Standard Model (MSM) which means that the particle
content of each sector contains no more than one Higgs doublet
and no right-handed neutrinos. This model has been discussed
previously (see Ref.\cite{flv}) with the motivation that the
discrete symmetry between the two sectors be interpreted as
an improper spacetime symmetry, e.g. parity. It is perhaps
interesting to note that this discrete symmetry remain intact
at low energies after gauge symmetry breaking.

What is of more interest is that this model could have testable
consequences at future collider experiments. This comes from
the non-gravitational interactions caused by the mixing of the
two sectors, viz., the mixing of the Higgs fields and the tree-level
gauge kinetic mixing term of the $U(1)$ fields\footnote{This is
\`a priori allowed if no assumption about gauge unification is made.}.
To establish notation, the interaction terms of interest are given
as follows: The Higgs potential is
\begin{equation}
V(\phi_1,\phi_2)  = \lambda_1\left(\phi_1^\dagger\phi_1
+ \phi_2^\dagger\phi_2 -2u^2 \right)^2
+ \lambda_2\left(\phi_1^\dagger\phi_1
- \phi_2^\dagger\phi_2  \right)^2.
\end{equation}
If $\lambda_{1,2} > 0$, then the minimum is the vacuum
\begin{equation}
\langle \phi_1 \rangle = \langle \phi_2 \rangle = (0, u)^T.
\end{equation}
As in the MSM, each Higgs doublet will give mass to the
$W^\pm$ and $Z$ bosons in their respective worlds after symmetry
breaking. There will also be two physical scalars, $\eta_{1,2}$,
which have non-degenerate masses due to the mixing. Their masses
are related by
\begin{equation}
{M_{\eta_1}\over M_{\eta_2}} = {\lambda_1 \over \lambda_2}
\equiv k.
\label{hm}
\end{equation}
The ordinary and shadow worlds can then communicate via the
Yukawa interactions mediated by the physical Higgs particles.
The Yukawa Lagrangian is given by
\begin{equation}
{\cal L}_{yuk} = {m\over \sqrt{2} u} \bar f f
\left( \eta_1 - \eta_2 \right)
+ {m\over \sqrt{2} u} \bar F F \left( \eta_1 + \eta_2 \right),
\label{yuk}
\end{equation}
where $f$ and $F$ are the generic ordinary and shadow fermions
respectively and $m$ is their mass. There will be two potentially
measurable effects if the masses of the two physical scalars are
quite different from each other. Firstly, the coupling of any one
of these Higgs particles to ordinary fermions is $(1/ \sqrt{2})$
that of the standard model Higgs boson. Therefore the probability
of producing one of these Higgs particles is $1/2$ that of the
standard model Higgs boson. Secondly, once one of these Higgs particles
has been produced then it will decay half of the time into
undetectable shadow matter and thereby the invisible decay
channels will make up half of its width.

The gauge kinetic interaction is given by
\begin{equation}
-{\cal L}_{KE} = {2\kappa\over 4} F_{1\mu\nu} F_2^{\mu\nu},
\end{equation}
where $F_{1,2}^{\mu\nu} =
\partial^\mu A_{1,2}^\nu - \partial^\nu A_{1,2}^\mu $
and $A_{1,2}^\mu$ are the $U(1)_{1,2}$ gauge fields respectively.
A rescaling and a rotation of the $U(1)$ fields puts the
gauge-fermion Lagrangian into standard form and results in a
modification of the electromagnetic couplings of the fermions.
The ordinary photon couples to
\begin{equation}
(Q_1 - \kappa \cos^2\theta_W Q_2)e
\end{equation}
while the shadow photon couples to
\begin{equation}
Q_2 e\sqrt{1 - \kappa^2\cos^4\theta_W},
\end{equation}
where $Q_{1,2}$ are the electric charge generators for the
ordinary and shadow worlds respectively (e.g. $Q_1 \not= 0$
and $Q_2 = 0$ for ordinary matter and vice versa for shadow
matter), $\theta_W$ is the usual electroweak mixing angle
and $e$ is the electromagnetic coupling. Therefore the charged
shadow particles will appear as mini-charged particles in the
ordinary world.\cite{mcp}

Having reviewed the possible experimental consequences of this
model, it would be interesting to see if the model is compatible
with standard big bang cosmology. Note that if the mixing
parameters between the ordinary and shadow worlds are made
sufficiently small, then the two sectors decouple, leaving
gravity as the only interaction between them. The cosmological
and astrophysical implications of such a model have been
discussed in Ref.\cite{kst}. It was found that the nucleosynthesis
bounds on the Helium abundance ruled out models which have exact
microscopic and macroscopic symmetries between the ordinary and
shadow worlds. However, it is still possible to have a model
with identical microphysics between the two sectors but with
different macrophysics. This macroscopic asymmetry can be quantified
by a temperature difference. Nucleosynthesis bounds require
$T_s/T \leq 0.68$ for three light neutrino species, where $T_s$
and $T$ are the temperatures for the shadow and ordinary sectors
respectively. One could attribute this temperature difference to
some as yet unknown dynamics (e.g. inflation) occuring at some early
epoch of the evolution of the Universe. In any case, the temperature
difference can be treated as an effective initial condition.
Since there are additional non-gravitational interactions in
the model considered here, it is necessary to check that these
interactions do not bring the two sectors into thermal
equilibrium given the initial temperature difference.

First, consider the gauge interactions between the two sectors,
i.e. $f\bar f \longleftrightarrow F \bar F$ mediated by a
s-channel photon. The two sectors will not come into equilibrium
if the interaction rate is less than the expansion rate of the
Universe. An estimate of the interaction rate is given by
\begin{equation}
\Gamma = n \sigma v_{rel} \sim \kappa^2\cos^4\theta_W\alpha^2_{em} T,
\end{equation}
where $n \sim T^3$ is the fermion number density, $\sigma$ is the
scattering cross section, $v_{rel}$ is the relative velocity
between the colliding particles and $\alpha_{em}$ is the usual
electromagnetic fine structure constant. The expansion rate of the
Universe for the radiation dominated epoch is
\begin{equation}
H = 1.66\sqrt{g_*}{T^2\over M_{pl}},
\label{ex}
\end{equation}
where $g_*$ counts the total number of effectively massless
degrees of freedom and $M_{pl}$ is the Planck mass. For the
situation when $\Gamma /H \leq 1$ then
\begin{equation}
\kappa \lsim \left(T\over {\rm MeV} \right)^{1\over 2} \times 10^{-9}.
\end{equation}

Now consider the Higgs sector, but at temperatures above the
electroweak phase transition (EWPT). The two sectors can
communicate via
$\phi_1 \phi_1^\dagger \longleftrightarrow \phi_2 \phi_2^\dagger$
and their respective Yukawa interactions with fermions. By estimating
the ratio of the interaction rate to the expansion rate of the
Universe for the Higgs scattering process, i.e.
\begin{equation}
{\Gamma \over H} \sim \left(\lambda_1 - \lambda_2 \right)^2
\left({{\rm GeV}\over T}\right) \times 10^{19},
\end{equation}
one finds that $\vert \lambda_1 - \lambda_2 \vert \lsim 10^{-8}$
for $T \sim 10^2$ GeV if $\Gamma / H \lsim 1$.
This is a disappointing result from an experimental/collider
physics point of view since the Higgs particles are nearly
degenerate (see Eq.(\ref{hm})) and hence the Higgs sector becomes the
same as that in the standard model. Therefore it seems that the
desirability of having a sizeable mass splitting between $\eta_1$ and
$\eta_2$ means that the two sectors will eventually come into thermal
equilibrium given an initial temperature difference.

This, however, is not the whole story. Now consider what happens
below the EWPT. At temperatures below the EWPT the Higgs-Higgs
scattering interaction will freeze out if the lightest
Higgs mass is assumed to be of the order of 100 GeV.
The two sectors will now communicate via their Yukawa interactions
(Eq.(\ref{yuk})) since the physical Higgs particles couple to both worlds.
The interaction rate for $f\bar f \longleftrightarrow F \bar F$ is
(for $m_f \lsim T \lsim M_{\eta}$)
\begin{equation}
\Gamma = {3g\zeta(3) \over 128\pi^3}
\left(m_f \over u \right)^4 {(k^2 -1)^2 \over k^4}
{T^5 \over M_\eta^4},
\end{equation}
where $g$ is the internal degrees of freedom of the scattering
fermion, $\zeta(3) \simeq 1.2$, $m_f$ is the fermion mass
and $M_\eta$ is the lightest Higgs boson mass (see Eq.(\ref{hm})).
Then by using Eq.(\ref{ex}) and $\Gamma / H \lsim 1$ gives a
decoupling temperature,
\begin{equation}
T_D \simeq \left\{ {64\pi^3\over \zeta(3)}
{g_*^{1\over 2}\over g}\left({u\over m_f}\right)^4
{k^4 \over (k^2 -1)^2 }{M_\eta^4 \over M_{pl}} \right\}^{1\over 3}.
\end{equation}
For $u \sim 10^2$ GeV, $m_f \sim 1$ GeV, $M_\eta \sim 10^2$ GeV,
$g_*^{1\over 2} / g \gsim {\cal O}(1)$ and $k = 2$ gives a decoupling
temperature around 1 GeV. (Note that smaller fermion masses will
give a higher decoupling temperature. For heavy fermions, such as
the t-quark with mass of the order of 100 GeV, their interactions
will freeze out.) So for temperatures below about 1 GeV
the two sectors are no longer in chemical equilibrium and hence
decouple. It is perhaps fortuitous that the decoupling of the
two sectors occur before nucleosynthesis begins. However, the
temperature of the two worlds remain the same. Thus, there exists
the necessary but not sufficient conditions for a departure from
thermal equilibrium between the ordinary and shadow worlds. So is
there any way to generate the required temperature difference,
\begin{equation}
{\Delta T\over T} \equiv {(T - T_s)\over T} \gsim 0.32,
\end{equation}
to satisfy the nucleosynthesis bounds?

Presumably, since the departure from chemical equilibrium
occurs at a temperature not too far away from the quark-hadron
phase transition (QHPT), there might be some hope that this
phase transition could generate the required out-of-equilibrium
conditions. One scenario is to consider thermal fluctuations
at the QHPT generating the temperature difference between the
ordinary and shadow worlds.\footnote{ One can also consider
thermal fluctuations in the cosmic fluid away from any phase
transition. It turns out that the fluctuations are very tiny.}
The hope here is that, in the process of nucleating bubbles of
the hadron phase, the specific heat of the hadron bubble is
sufficiently small to generate a large enough temperature
difference between the two sectors. This scenario assumes a
first order QHPT \cite{lat91} and uses a simple model of
nucleation \cite{ll,others}. The details of this scenario can be
described as follows.

When the temperature falls below the critical value\footnote{
It is assumed that the critical temperature of the QHPT
is the same for the ordinary and shadow worlds.}, $T_c$,
the quark-gluon plasma is supercooled and bubbles of hadron
gas nucleate and grow. For the remainder of this paper it will
be assumed that the amount of supercooling will be relatively
small. The bubbles of the hadron phase grow by deflagration
heating of the surrounding plasma back to a temperature near
$T_c$ where the two phases coexist in unstable thermal equilibrium.
Eventually the hadron phase grows slowly at the expense of the
quark-gluon phase. When the hadron phase occupies about 50\% of the
space then the plasma begins to shrink. This is the qualitative
picture of what is believed to have happened during a first order
cosmological QHPT.

Now consider the stage when the hadron phase is nucleated. Only
bubbles of a certain minimum or critical size, $r_c$, survive
to grow and form a centre of condensation for the formation of
a new phase. Bubbles smaller than the critical size will shrink
and eventually disappear. The critical radius of the bubble is
given by
\begin{equation}
r_c = {2\gamma \over P_h - P_q },
\label{rc}
\end{equation}
where $\gamma$ is the surface tension between the two phases
and $P_h - P_q$ is the pressure difference between the two
phases. It is known \cite{ll} that the surface tension depends
on the temperature and vanishes at $T_c$. From dimensional
arguments, $\gamma \sim T_c^3 f\left({T\over T_c}\right)$,
where f is some unknown function. Without any additional
information, it might be reasonable, for small supercooling,
to expand $\gamma (T)$ near $T = T_c$
(assuming ${d\gamma \over dT}$ to be finite) giving
\begin{equation}
\gamma \simeq 3\gamma_0 T_c^2 \Delta T,
\label{st}
\end{equation}
where $\gamma_0$ is a dimensionless constant expected to
be of order one and $\Delta T = T_c - T$. ( Eq.(\ref{st}) is
equivalent to writing $\gamma (T)$ as
$\gamma_0 \left( T_c^3 - T^3 \right)$. )
The pressure difference can be estimated by assuming that the
relativistic degrees of freedom dominate the pressure: i.e.
only pions, u- and d-quarks, gluons and their shadow world
counterparts contribute to the pressure difference for
$T \sim T_c = 100 - 200$ MeV. This then gives
\begin{equation}
P_h - P_q =
6\left({\pi^2\over 90}\right) T^4
-74\left({\pi^2\over 90}\right) T^4 + P_{vac},
\end{equation}
where $P_{vac}$ is the pressure of the nonperturbative vacuum
exerted on the perturbative vacuum. $P_{vac}$ can be written
in terms of the critical temperature, which is defined as the
temperature when $P_h = P_q$.
Thus $P_{vac} = 68\left(\pi^2 / 90\right) T_c^4$. For $T$ near
$T_c$,
\begin{equation}
P_h - P_q \simeq
4\times 68\left({\pi^2\over 90}\right) T_c^3 \Delta T.
\end{equation}
Hence the critical radius of the bubble (see Eq.(\ref{rc})) is
\begin{equation}
r_c = { 3\gamma_0 \over 2\times 68 }\left({90\over \pi^2}\right)
{1\over T_c}.
\label{rc2}
\end{equation}
The specific heat of the bubble can now be estimated to give
\begin{eqnarray}
C_V & = & {d({\rm Energy})\over dT} \quad \hbox{ at constant volume}
\nonumber  \\
 & \simeq & 2\times 12\left({\pi^2\over 30}\right)T^3
\left\{ {4\pi\over 3} r_c^3 \right\},
\end{eqnarray}
where the term in the last set of parenthesis is the volume
of the bubble. Substituting Eq.(\ref{rc2}) into the above gives
\begin{equation}
C_V \sim 12\pi \left({3\gamma_0\over 68}\right)^3
\left({90\over \pi^2}\right)^2
\left({T\over T_c}\right)^3.
\end{equation}
This result in turn allows an estimation of the size of
the temperature fluctuations in the hadron bubble. If the
fluctuations are not too large in size then they can be
described by a Gaussian distribution. Their root-mean-squared
value (RMS) is
\begin{equation}
\left(\Delta T \right)_{RMS} = {T\over\sqrt{C_V}}
\sim 2\gamma_0^{-{3\over 2}}T_c
\end{equation}
for $T \sim T_c$. By requiring $0.32 < \left(\Delta T \right)_{RMS}/T < 1$
gives $1.6 \lsim \gamma_o \lsim 3.4$ which is consistent with the
order of one expectations for the value of $\gamma_0$.
Therefore it appears plausible that fluctuations during the nucleation
of the hadron phase can produce the required temperature difference
to be established between the ordinary and shadow worlds.

To elucidate the above discussion, the proposed scenario can summarised
as follows: Nucleosynthesis requires that the ordinary and shadow
worlds be at different temperatures given identical microphysics
between the two worlds. For $\kappa \lsim 10^{-9}$, the electromagnetic
interaction will not contribute to bringing the two worlds into
thermal equilibrium given an initial temperature difference between
the two worlds. The Higgs interaction, for an experimentally interesting
paramter space, will be of sufficient strength to bring the two sectors
into equilibrium at temperatures above the EWPT. Below the EWPT the
two sectors are no longer in chemical equilibrium when the temperature
is around 1 GeV. However, the two sectors will still be at the same
temperature. It was conjectured that the nucleation of the
hadron phase at the QHPT can produce large enough fluctuations
to establish the required temperature difference between the
two worlds.

So far it has been shown that the model with an experimentally
interesting parameter space could be compatible with standard
big bang cosmology. It might be pertinent to check if there
are any contraints from astrophysical considerations. An obvious
constraint comes from the possible energy loss of a neutron star
when ordinary particles in the star can produce shadow particles
via the electromagnetic and Higgs interactions. Since the shadow
particles interact very weakly with the ordinary world then there
exists the possibility that the shadow matter carries off too
much energy and cools the star too rapidly. An estimate of this
energy loss is given by
\begin{equation}
Q = V_{core} n^2 \sigma v_{rel} \langle E \rangle,
\end{equation}
where $V_{core}$ is the volume of the core of the star,
$n = \left(3g\zeta (3)/4\pi^2\right)T^3$ is the number density,
$\sigma$ is the interaction cross section, $v_{rel}$ is the
relative velocity between the interacting particles and
$\langle E \rangle = \left( 7\pi^4/ 180\zeta (3)\right) T$
is the average energy of the particles in the star. For core
temperatures of the order of 10 MeV, it is expected that only
electrons participate in the production of shadow matter. Taking
the core radius to be 10 km, the energy loss for the electromagnetic
interaction is $Q \sim 10^{66}\kappa^2$ erg/s, while for the
Higgs interaction, it is $Q \sim 10^{32}$ erg/s. By requiring that
the energy loss be less than $10^{53}$ erg/s gives
$\kappa \lsim 10^{-7}$. Therefore astrophysics gives a relatively
weak constraint on the shadow model parameters.

In conclusion, it has been shown that the existence of a shadow
matter world with identical microphysics to that of the
ordinary world can have an interesting experimental parameter space
and at the same time need not be incompatible with standard big bang
cosmology if the macrophysics of the two worlds differ. It was
speculated that the macro-asymmetry between the two sectors
can be generated by fluctuations at the quark-hadron phase transition.

\vspace{5 mm}
\centerline{\bf Acknowledgements}
This work was supported in part by a grant from the DOE.
The author wishes to thank Robert Foot, T.K. Kuo, Paul Muzikar
and Ray Volkas for discussions.

\end{document}